\documentclass{ws-procs975x65}

\newcommand{\be}{\begin{eqnarray}}
\newcommand{\ee}{\end{eqnarray}}
\def\lsim{\mathrel{\rlap{\lower3pt\hbox{\hskip1pt$\sim$}}
     \raise1pt\hbox{$<$}}} 
\def\gsim{\mathrel{\rlap{\lower3pt\hbox{\hskip1pt$\sim$}}
     \raise1pt\hbox{$>$}}} 

\def\la{\langle}\def\ra{\rangle}
\def\del{\partial}
\def\tr{{\rm tr}}
\def\calL{\cal L}

\newcommand\Tr{\mathrm{Tr}\,}

\def\bi{\bibitem}

\begin{document}

\title{Cold Compressed Baryonic Matter \\ with Hidden Local Symmetry and Holography}

\author{Mannque Rho$^*$}

\address{ Institut de Physique Th\'eorique,  CEA Saclay, 91191 Gif-sur-Yvette C\'edex, France\\
 and Department of Physics, Hanyang University, 133-791 Seoul, Korea\\
$^*$E-mail: mannque.rho@cea.fr\\ http://ipht.cea.fr, //hadron.hanyang.ac.kr}

\begin{abstract}
I describe a novel phase structure of cold dense baryonic matter predicted in a hidden local symmetry approach anchored on gauge theory and in a holographic dual approach based on the Sakai-Sugimoto model of string theory. This new phase is populated with baryons with half-instanton quantum number in the gravity sector which is dual to half-skyrmion in gauge sector in which chiral symmetry is restored while light-quark hadrons are in the color-confined phase. It is suggested that such a phase that aries at a density above that of normal nuclear matter and below or at the chiral restoration point can have a drastic influence on the properties of hadrons at high density, in particular on short-distance interactions between nucleons, e.g., multi-body forces at short distance and hadrons -- in particular kaons -- propagating in a dense medium. Potentially important consequences on the structure of compact stars will be predicted.
\end{abstract}


\bodymatter

\section{The Issue}\label{sec1}
Baryonic matter near the nuclear matter density $n_0\approx 0.16$ fm$^{-3}$ is very well understood, thanks to many years of nuclear experimental and theoretical efforts, but there is a dearth of experimental information beyond $n_0$. Unlike in high temperature where lattice QCD calculations come in tandem with experiments performed at relativistic heavy-ion colliders, there are no reliable theoretical tools available for dense matter for which lattice gauge calculation suffers from the notorious sign problem. Thus beyond $n_0$, one knows very little of what's going on. At asymptotically high density, perturbative QCD with weak coupling allows a clear-cut prediction, but the density required there is so high that it is hardly likely to be relevant to the physics of dense matter in terrestrial laboratories or in compact stars. A large number of model calculations are nonetheless available in the literature with a plethora of predictions for nucleon stars, quark stars etc., but the problem with these predictions is that constrained or aided neither by experiments nor by theory, it is difficult to assess the reliability of the calculations with wildly varying results at densities going beyond that of the nuclear matter.

The prospect for the future, however, is pretty good, in particular experimentally. Indeed the forthcoming accelerators devoted to the physics of dense matter, such as FAIR/GSI, will help build realistic phenomenological models just like the ones in standard nuclear physics where the wealth of accurate data played a pivotal role in arriving at realistic nuclear structure models with no guidance from the fundamental theory of strong interactions QCD. It is now established, thanks to the recent development in QCD-based effective field theories of nuclei and nuclear matter, that what the nuclear physicists have been doing was on the right track all along. While awaiting the advent of experimental inputs for going toward dense matter, the challenge for the theorists is: Make predictions, relying solely on theoretical schemes that are as faithful as possible to QCD proper, for the poorly understood regime between the normal nuclear matter density $n_0$ and the chiral transition density $n_\chi$ -- at which the restoration of chiral symmetry is expected to take place. In this talk, I would like to describe one such (theoretical) effort being carried out in the context of the World Class University (WCU) Program at Hanyang University in Seoul, Korea, that is anchored on the exploitation of the notion of hidden local symmetry in strong interactions and its holographic generalization following from gauge/string duality. What I will describe is just the beginning. As such, what's predicted at present cannot be immediately confronted with nature. But it will be easily subjected to confirmation or falsification when the data become available.

This talk is not strictly in the main line of this conference where the principal focus is on strong coupling phenomena in the LHC era. My talk concerns  rather physics under extreme conditions at a much lower energy scale which will be probed at other laboratories. The basic concept involved, however, such as hidden local symmetry and gauge/string duality, turns out to be closely related to the main theme of this conference.
\section{Hidden Local Symmetry (HLS)}
\subsection{HLS Lagrangian}
Hidden local symmetry (HLS) in the strong interactions\footnote{Here I will be focusing mainly on the two-flavor (u,d) case in the chiral limit, i.e., with zero quark mass, but will introduce the strangeness (s) quantum number as needed.}  is a flavor gauge symmetry, the effective degrees of which consist of Goldstone bosons (or more realistically pseudo-Goldstone bosons in Nature), i.e., the pions ($\pi=\frac 12 \vec{\tau}\cdot \vec{\pi}$) and the vector fields $V_\mu$ valued in $U(2)$ (i.e., $V_\mu=\frac 12\vec{\tau}\cdot\vec{\rho}_\mu +\frac 12 \omega_\mu$), which elevates the energy scale of the chiral symmetric interactions from the current algebra scale with soft pions to the vector-meson scale $m_V\sim 800$ MeV.

There are two ways to view how hidden gauge symmetry in the strong interactions arises. One is to view the local symmetry as an ``emergent" symmetry that exploits the redundancy of local gauge symmetry and the other is to view it as descending (via Kaluza-Klein reduction) from higher dimensions.

In Nature, in addition to the pions and the vector mesons $V_\mu$, there are other light mesonic excitations such as scalar mesons and axial-vector mesons. As for the baryons -- that are indispensable for dense matter, they will be generated as solitons from the HLS Lagrangian, instead of positing them as is done in standard treatments. This approach is most natural from the point of view of large $N_c$ QCD.

Now limited to the minimum number of degrees of freedom, the gauge-invariant Lagrangian takes the simple form\footnote{In what follows, the parametric pion decay constant will be denoted $F_\pi$ while the physical pion decay constant will be written as $f_\pi$.}
\be
{\calL}= \frac{F_\pi^2}{2}{\Tr}\{|D_\mu\xi_L|^2 + |D_\mu\xi_R|^2
 + \frac{\gamma}{2} |D_\mu U|^2\} \ - \frac{1}{2} \,
\mbox{Tr} \left[ V_{\mu\nu} V^{\mu\nu} \right] +\cdots \label{hls}
 \ee
where $U=e^{2i\pi/F_\pi}=\xi^\dagger_L\xi_R$, $\xi_L=e^{-i\frac{\pi}{F_\pi}}e^{i\frac{\sigma}{F_\sigma}}$, $\xi_R=e^{i\frac{\pi}{F_\pi}}e^{i\frac{\sigma}{F_\sigma}}$, $D_\mu\xi$ is the covariant derivative, and $\gamma=1/a-1$ with $a\equiv F_\sigma^2/F_\pi^2$. In this form, this is an elegant, though highly nonlinear, Lagrangian. However being an effective Lagrangian defined with a cutoff, higher derivative terms at loop orders indicated by the ellipsis in (\ref{hls}) must figure at the quantum level.

In matter-free space, namely, at zero temperature ($T=0$) and zero density ($n=0$), the HLS Lagrangian is gauge-equivalent to the standard and well-tested chiral Lagrangian with the pions alone, and as such leads to the same chiral perturbative procedure for pion interactions at very low energy as the standard (pion-only) chiral Lagrangian\cite{HY:PR}. At tree order, the hidden local symmetry is of no special power or advantage; gauge fixed to unitary gauge, it will simply reproduce the current algebra results of the standard chiral Lagrangian. The local symmetry, however, picks up its power at higher loop orders since it then renders -- in principle -- feasible a systematic power counting including the vector mesons, a task that is highly  cumbersome if not impossible in a gauge-fixed theory\cite{georgi-theoryspace}. What is perhaps not properly appreciated among nuclear theorists investigating hadronic matter under extreme conditions (at high $T$ and/or high $n$) is that hidden local symmetry can incorporate {\em with ease} certain properties of hadronic matter that involve highly correlated many-body interactions that are difficult to access by models that do not possess flavor local symmetry. The purpose of this talk is to discuss what transpires when HLS in its simplest form and in certain approximations is applied to cold dense matter beyond the nuclear matter density. I will present certain intriguing predictions that differ from standard phenomenological many-body models found in the literature.

It has been shown at one-loop order in matter-free space that the HLS theory matched to QCD (in terms of correlators) flows to a unique fixed point called ``vector manifestation" when temperature or density approaches the critical point where chiral restoration takes place\cite{HY:PR}\footnote{It is easy to see  that this statement should hold to higher loop orders although there is no mathematically rigorous proof. In practice, higher orders are not very useful at present in view of the large number of unknown parameters that arise at higher orders. This means that the results I will show cannot be taken to be quantitatively accurate, even if qualitatively sound.}. In the presence of dense matter, the parameters of the Lagrangian governed by  renormalization group equations flow as density increases.\footnote{Unless specifically mentioned otherwise, similar arguments apply to temperature. In contrast to temperature, to take into account the density, fermion degrees of freedom, here baryons, need be introduced. This problem will be addressed below.} When matched to the QCD vector and axial vector correlators via the in-medium OPE, that is, with the quark condensate $\la\bar{q}q\ra$ and the gluon condensate $\la G^2\ra$ endowed with background density dependence, the parameters of the HLS Lagrangian (\ref{hls}) tend, as density approaches the chiral transition density $n_\chi$, to
\be
g^\star \propto  \la\bar{q}q\ra^\star\rightarrow 0,\ a^\star \equiv  F^\star_\sigma/F^\star_\pi\rightarrow 1\label{VM}
\ee
where the asterisk stands for density dependence. In contrast, there turns out to be nothing special with the flow of the decay constants $F^\star_{\sigma,\pi}$ for varying densities although the physical decay constant $f^\star_\pi$, which receives radiative corrections, should go to zero as $\la\bar{q}q\ra^\star$ goes to zero. The consequence of the VM point (\ref{VM}) is that the parametric mass $m^\star_V$ vanishes
\be
m^\star_V\approx a^\star F^\star_\pi g^\star\propto g^\star \rightarrow 0.\label{MV}
\ee
Equations (\ref{VM}) and (\ref{MV}) are the essential ingredients of the hidden local symmetry theory with the VM -- denoted ``HLS/VM" for short -- that I will resort to in what follows.

Now the important practical question is: Where and how are these fixed point values manifested? In other words, how can one ``see" the presence of these fixed points that are the potential signals for the chiral symmetry structure of the hadronic vacuum? This represents one key question  nowadays asked by nuclear physicists.

To answer this question, one should recognize the caveats in applying this Lagrangian to Nature. First of all, the Lagrangian itself is a highly truncated one, with the minimum number of relevant degrees of freedom, so cannot be expected to work accurately for certain processes that require other degrees of freedom. Secondly, to confront experimental observables, dense loop corrections must be calculated in conformity with chiral perturbation theory, what is most important, with the parameters of the Lagrangian {\em running with density} and satisfying thermodynamic and symmetry constraints. This means that the signals of HLS/VM cannot in general be singled out in a simple manner. They are likely to be compounded with density dependence brought in by dense loop corrections. As in the case of high-temperature processes in relativistic heavy-ion collisions, e.g., dilepton production, a careful weeding-out of ``trivial effects" of many-body nature will be required before one can see the desired effects. This issue is discussed in detail in ~\cite{BHHRS}.
\subsection{Gauge-fixed Lagrangian}
I will take the Lagrangian (\ref{hls}) in unitary gauge. It has the form
\begin{eqnarray}
{\cal L}_{hls} &=& \frac{f_\pi^2}{4}
\mbox{Tr}(\partial_\mu U^\dagger \partial^\mu U)
-\frac{f_\pi^2}{4} a
 \mbox{Tr}[\ell_\mu + r_\mu + i(g/2)
( \vec{\tau}\cdot\vec{\rho}_\mu + \omega_\mu)]^2\nonumber\\
&& -\textstyle \frac{1}{4} \displaystyle
\vec{\rho}_{\mu\nu} \cdot \vec{\rho}^{\mu\nu}
-\textstyle \frac{1}{4}  \omega_{\mu\nu} \omega^{\mu\nu}
+\cdots,
\label{lag}\end{eqnarray}
with $
U =\exp(i\vec{\tau}\cdot\vec{\pi}/f_\pi) \equiv \xi^2$,
$
\ell_\mu = \xi^\dagger \partial_\mu \xi, \mbox { and }
r_\mu = \xi \partial_\mu \xi^\dagger$,
$
\vec{\rho}_{\mu\nu} = \partial_\mu \vec{\rho}_\nu
- \partial_\nu \vec{\rho}_\mu + g \vec{\rho}_\mu \times \vec{\rho}_\nu$ and
$\omega_{\mu\nu}=\partial_\mu\omega_\nu-\partial_\nu\omega_\mu$. Because of the presence of the isoscalar vector meson $\omega$, there is an additional Lagrangian which is parity-odd
\be
{\cal L}_{hWZ}=
+\textstyle\frac{3}{2} g \omega_\mu B^\mu\label{hwzterm}
\ee
with the baryon current
\be
B^\mu =  \frac{1}{24\pi^2} \varepsilon^{\mu\nu\alpha\beta}
\mbox{Tr}(U^\dagger\partial_\nu U U^\dagger\partial_\alpha U
U^\dagger\partial_\beta U).
\ee
This is a part of the gauged Wess-Zumino term that survives for $N_f <3$ while the topological 5D Wess-Zumino term is absent in the two-flavor case I am discussing. In what follows, it will be called  ``hWZ term" -- ``h" standing for homogenous. In general the hWZ Lagrangian has four (homogeneous) terms with arbitrary coefficients but if one imposes vector dominance in the isoscalar channel and consider the $\rho$ meson to be so massive that one can substitute $\rho_\mu^a=i{\Tr}[\tau^a(l_\mu + r_\mu)]$, then it reduces to one term of the form (\ref{hwzterm})\cite{meissner}.\footnote{There are two caveats in this. One is that vector dominance is most likely violated with $a$ approaching 1 in hot and dense matter\cite{HS:VD}. The second is that the ``large $\rho$ mass limit" is at odds with the VM where the vector meson mass goes to zero. These two caveats can be avoided if one works with all the hWZ terms which make the calculation considerably more involved. The merit of the form (\ref{hwzterm}) is that it is consistent with the Chern-Simons term in 5D that I will consider in the holographic approach discussed below.}
\subsection{Describing dense matter}
The Lagrangian ${\calL}={\cal L}_{hls}+{\cal L}_{hWZ}$ contains no {\em explicit} baryon degrees of freedom. However there is a topological baryon current $B_\mu$ in (\ref{hwzterm}), so one can think of the $\omega$ field as a chemical potential conjugate to the baryon charge. I will parallel this structure below to a holographic dual form in 5D where a similar structure arises.

One might attempt to describe dense baryonic matter with (\ref{lag}) and (\ref{hwzterm}) in the mean field. However this approach simply does not work in the absence of explicit baryon degrees of freedom. One natural way is to generate skyrmions as solitons from the Lagrangian. In principle, there can be soliton solutions with baryon number going from 1 to infinity, the latter corresponding to nuclear matter. Indeed, there have been attempts to build finite nuclei with mass number up to, say, $\sim 27$ which would correspond to a skyrmion with baryon charge $B=27$\cite{manton}. The resulting theory with skyrmions coupled to fluctuating vector and pion fields, properly quantized, could be expressed in terms of local baryon fields coupled to the mesons satisfying the same symmetry structure as the starting theory. This is in the same spirit as the baryon chiral Lagrangian where local baryon fields are coupled to pions in a chirally invariant away, which is the basis of chiral perturbation theory for pion-nucleon interactions. Given such a baryon chiral Lagrangian, then one can do a mean field calculation for many-body systems that would correspond to what is called Walecka mean field theory in nuclear physics. This strategy, backed by experimental data, can be found to be successful up to the density for which data are available. But how to apply it to a matter beyond the nuclear matter density is not known since there have been no experiments that probed the given high density regime. This is because at higher densities, higher dimension operators must enter, which  correspond, in the localized form, to rendering the constants of the mean field Lagrangian density-dependent and the density dependence is not known by theory alone.

What is in principle more powerful and closer to QCD proper is to treat many-body systems in terms of the solitonic solution of the Lagrangian. The soliton of winding number $A$ is the solution for an $A$-body baryonic system. The system with $A=\infty$ will then be a nuclear matter. Nobody has been able to perform an analytic or continuum calculation of such multi-body systems but there have been a variety of calculations putting the skyrmions on a crystal lattice. At high density, one expects the system to be in a crystal and so this is a natural approach to the problem although at lower density nuclear matter is known to be in a liquid than in a crystal. In fact, it has been established that nearly independently of detailed structure of the Lagrangian, there is a phase change from a matter of skyrmions at low density to a matter consisting of half-skyrmions at higher density\cite{halfskyrmions}, and this phase change takes place almost independently of whether the Lagrangian contains local gauge fields or not. The resulting half-skyrmion phase possesses certain scale invariance. What could be different is the critical density at which this takes place, which depends on the parameters of the Lagrangian as well as the degrees of freedom involved.

The crystal structure of dense skyrmion matter with a variety of different ansatze for the skyrmoion configuration is reviewed recently by Park and Vento\cite{park-vento}. For latter purpose, I will restrict myself to the approach\cite{prv-atiyah-manton} based on the Atiyah-Manton ansatz\cite{atiyah-manton}, since it is closely connected to the instanton picture that arises in 5D Yang-Mills Lagrangian of holographic dual QCD.

The Atiyah-Manton ansatz for the chiral field $U$ is given by the holonomy in (Euclidean) time direction
\be
U(\vec{x})=CS\left\{P{\rm exp}\left[-\int^\infty_{-\infty} A_4 (\vec{x},t)dr\right]\right\}C^\dagger\label{AM}
\ee
where $A_4$ is the (Euclidean) time component of an instanton field of charge $B$ -- which is $A$ for the $A$-baryon system, $P$ is the path ordering, $S$ is a constant matrix to make $U$ approach 1 at infinity and $C$ denotes an overall $SU(2)$ rotation. The homotopy assures that the static soliton configuration carries the same baryon number as the total charge of the instanton. Note that here YM gauge field $A_\mu$ is a fictitious field, not present in the theory (\ref{hls}). In holographic QCD described with a 5D YM field, $A_4$ is the gauge field present in the 5th direction in the theory.

\begin{figure}[h]
\centerline{\psfig{file=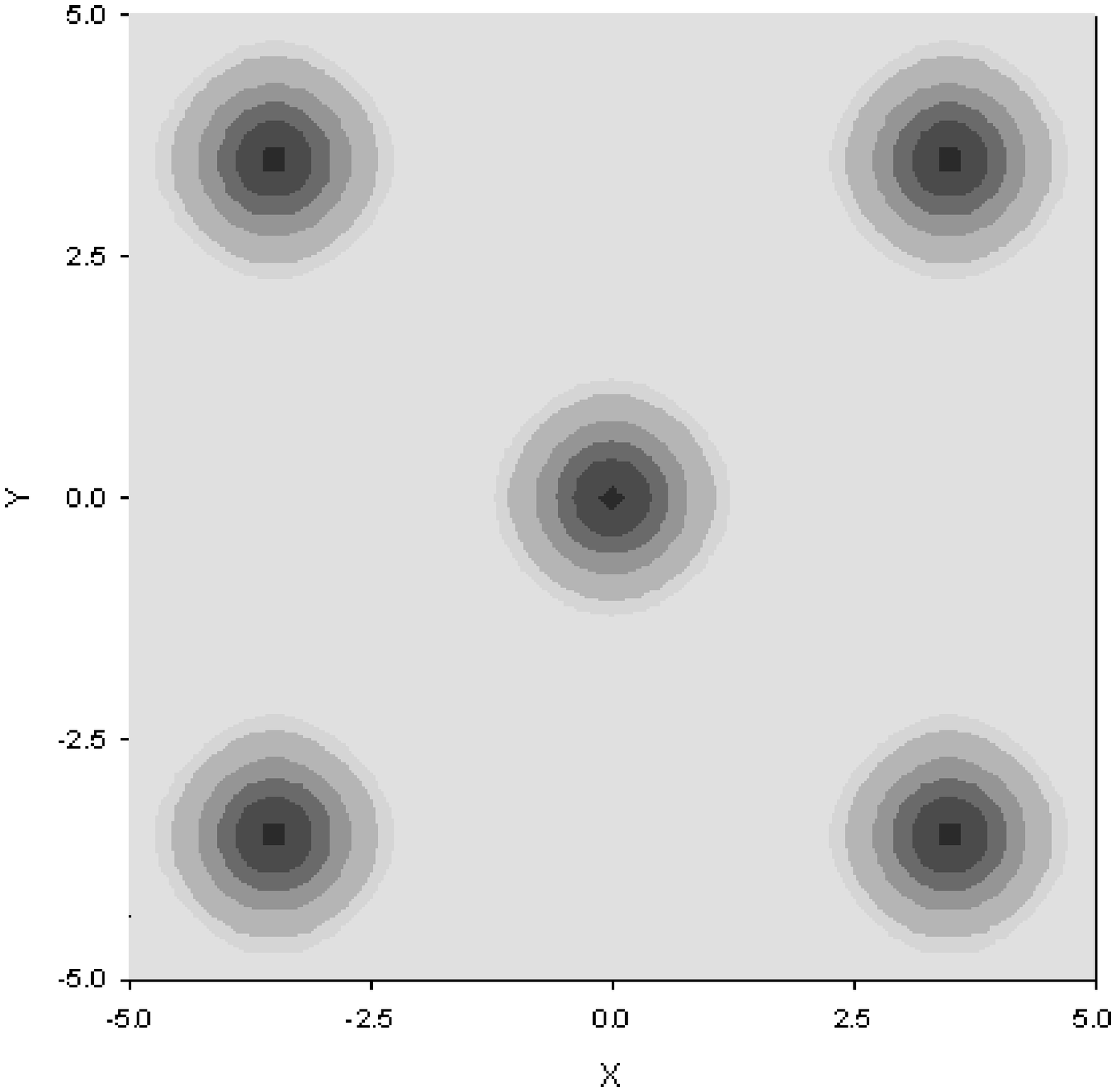,width=6cm,angle=0}
\psfig{file=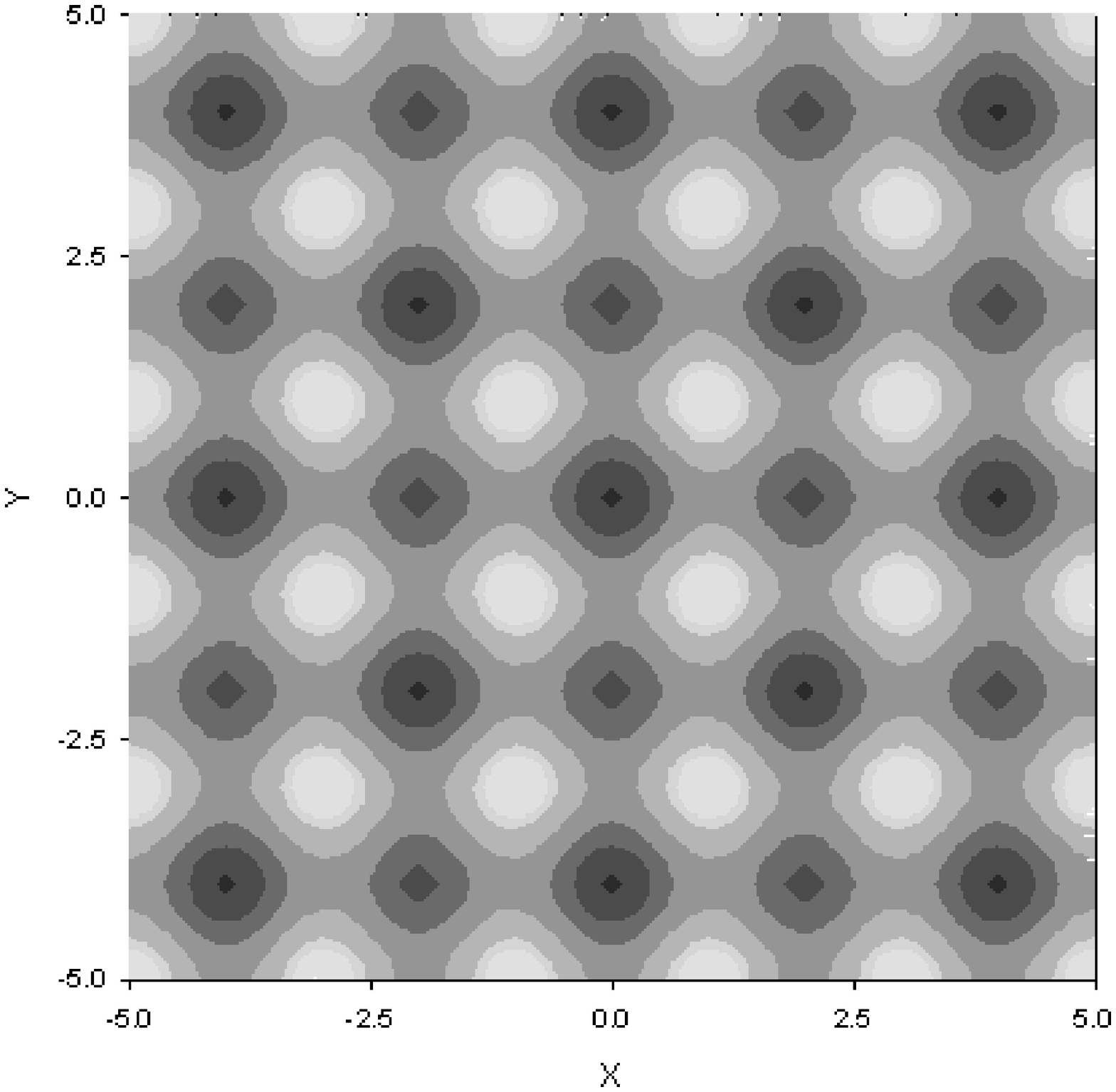,width=6cm,angle=0}}
 \caption{Local baryon
number densities at $L=3.5$ and $L=2.0$. For
$L=2.0$ the system is a half-skyrmion (or nearly half-skyrmion with mass) in a CC crystal configuration.}
\label{density}
\end{figure}

In \cite{prv-atiyah-manton}, the ansatz (\ref{AM}) is employed in putting $A$ instantons in an FCC crystal to compute the ground state energies as a function of the crystal lattice size $L$\footnote{In FCC, the baryon number density is given by $n=1/2L^3$.}. By shrinking the size $L$, the multi-skyrmion system is {\em by fiat} compressed. In Nature, gravity does the job of squeezing. In Fig.~\ref{density} is shown the local baryon number density profile before and after the splitting. These results are obtained with the Skyrme Lagrangian with massive pions but the generic features are the same with the HLS Lagrangian (\ref{hls}). What is important with this phase change is that the half-skyrmion phase has $\la\bar{q}q\ra\propto {\rm Tr}U= 0$ but $f_\pi\neq 0$ whereas the skyrmion phase has $\la\bar{q}q\ra\neq 0$ and $f_\pi\neq 0$, which is symptomatic of chiral symmetry in the Nambu-Goldstone mode. What is novel here is that the chiral symmetry is ``restored" whereas there is propagating pion with non-vanishing decay constant. This means that quarks are still confined in that phase although chiral symmetry is ``restored." This feature is quite robust.
This result implies that in the half-skyrmion phase, the nucleon mass going roughly like $f_\pi^\star$ will not drop appreciably as does the vector meson mass. This will have an important consequence on nuclear physics at high density.

I should mention that a similar half-skyrmion phase structure is found in condensed matter physics (in 3D)\cite{condensed}. There the half-skyrmions get deconfined, that is, they are unbound. In the strong-interaction case at hand, one cannot say whether the half-skyrmions are bound or not~\footnote{In holographic QCD, it appears that the half-instantons are bound although maximally separated by Coulomb repulsion\cite{RSZ}.}.

\subsection{Meson fluctuations and dilatons}
Given the background with skyrmions and half-skyrmions, the interesting question is how mesons behave in dense medium.  This question has been addressed with the Lagrangian given by Eqs.~(\ref{lag}) and (\ref{hwzterm}). What is found there is illuminating: It indicates that without additional degree(s) of freedom, the VM property of HLS fails to manifest with the unitary gauge Lagrangian in medium\cite{prv-vector}. It is found that a dilaton field is rquired\cite{prv-dilaton}.

What happens to meson properties in dense medium is crucially controlled by the hWZ term (\ref{hwzterm}). In mean field in medium, it contributes
\be
{\cal L}_{hWZ}=
+\textstyle\frac{3}{2} g \omega_0 (x) n (x) \label{wzterm1}
\ee
where $n(x)$ is the nuclear matter distribution. Now the $\omega$ meson gives rise to a Coulomb potential, so the hWZ term leads to a repulsive interaction.
%
%
What is most important is that this repulsive interaction turns out to dominate over other terms, and the repulsion diverges as density increases. In order to prevent the energy per baryon coming from the hWZ term from diverging, $m_\omega^*$ has to increase sufficiently fast. And since $m_\omega^*\sim f_\pi^* g$ in this model, for a fixed $g$, $f_\pi^*$ must therefore {\em increase}. This is at variance with Nature: QCD predicts that it should {\em decrease} and go to zero -- in the chiral limit -- at the chiral transition. The problem here is that the VM property -- that $g\rightarrow 0$ as density approaches the critical -- is missing in this gauge-fixed mean field approach. Stated concisely, the presence of the $\omega$ field prevents both the pion decay constant $f_\pi$ and the $\omega$ mass from decreasing as demanded by the HLS/VM. The quark condensate $\la\bar{q}q\ra$ does ultimately vanish at a critical density but with an increasing $f_\pi$. The $\omega$ field thus brings havoc in the theory.

This defect can be remedied by the dilaton field associated with the trace anomaly of QCD\cite{prv-dilaton}\footnote{There is no elegant and simple way known of introducing scalar fields in the HLS theory. What I describe here may not be the most efficient way for HLS {\em per se} but it is one simple and elegant way for an effective field theory. The role of the dilaton field $\chi_S$ in what is referred to as Brown-Rho scaling\cite{br91} has been incorrectly understood by numerous authors who have argued against it on theoretical grounds.

Although it is not in the main line of discussions of this talk, it is worth mentioning here that the dilaton introduced therewith resolves one embarrassing feature of the Skyrme model for the nucleon, that is that, with the parameters of the Lagrangian given by the meson sector, the soliton mass comes out much too high, say, $\sim 1.5$ GeV. When the dilaton is put with a relatively low mass, say, $\lsim 1$ Gev, then the soliton mass drops to near the physical nucleon mass. The scalar degree of freedom carries a strong attraction that cannot be ignored in the single baryon as well as  baryonic matter structure.}.
In \cite{prv-dilaton}, two-component scalar fields denoted as ``soft" $\chi_S$ and ``hard" $\chi_H$ satisfying the QCD trace anomaly were introduced. Roughly speaking the soft component represents the scalar that is locked to the quark condensate, hence directly connected to chiral symmetry, while the hard component represents the glue-ball degrees of freedom tied to the scale-invariance breaking associated with the asymptotic running of the color coupling constant $g_c$. Given that $\chi_H$ involves high-energy excitations at a scale much greater than that of $\chi_S$, it can be integrated out for our purpose. Thus retaining only the $\chi_S$ field, one can modify Eqs.~(\ref{hls}) and (\ref{hwzterm}) so that the trace anomaly is suitably implemented. How this can be done in consistency with the QCD trace anomaly is described in\cite{HKL-MR}.

With the dilaton included consistently with the scale anomaly, the effective HLS Lagrangian in unitary gauge is
\be
{\cal L}={\cal L}_{hls\chi}+ +{\cal L}_{hWZ}\label{lagtot}
\ee
where
\be
{\cal L}_{HLS\chi} &=& \frac{f_\pi^2}{4}(\chi/f_\chi)^2
\mbox{Tr}(\partial_\mu U^\dagger \partial^\mu U)
-\frac{f_\pi^2}{4} a(\chi/f_\chi)^2
 \mbox{Tr}[\ell_\mu + r_\mu + i(g/2)
( \vec{\tau}\cdot\vec{\rho}_\mu + \omega_\mu)]^2\nonumber\\
&& -\textstyle \frac{1}{4} \displaystyle
\vec{\rho}_{\mu\nu} \cdot \vec{\rho}^{\mu\nu}
-\textstyle \frac{1}{4}  \omega_{\mu\nu} \omega^{\mu\nu}
+\frac 12 \del_\mu\chi\del^\mu\chi + V(\chi)\label{lags}\\
{\cal L}_{hWZ} &=& \textstyle\frac{3}{2} g (\chi/f_\chi)^n \omega_\mu B^\mu\label{fwzterm}
\ee
with $\chi\equiv\chi_S$ where $f_\chi$ is the $\chi$ decay constant and $V(\chi)=B\chi^4{\rm ln}\frac{\chi}{f_{\chi}e^{1/4}}$
is the potential that gives the energy-momentum tensor whose trace gives the soft component of the trace anomaly. The exponent $n$ which cannot be fixed by first principles can be arbitrary with $n>1$ without violating scale symmetry\cite{HKL-MR}. In the analysis of \cite{prv-dilaton}, $n=3$ was taken for simplicity, but $n=2$ gives an equally satisfying result at least qualitatively.

The lesson from this model is that the dilaton field plays a crucial role in making HLS theory compatible with nature. It makes both the $\omega$ meson mass and the pion decay constant decrease as density increases in consistency with HLS/VM. An additional important consequence of the dilaton is that it reveals a novel phase with half-skyrmions as the relevant degrees of freedom where the quark condensate vanishes, $\la\bar{q}q\ra=0$, while $f_\pi\neq 0$. This phase appears at a density $n_{1/2}$ above the nuclear matter density $n_0$ but below the chiral transition density $n_\chi$. This phase is characterized by the restored chiral symmetry but with quarks {\em confined}. At what density this phase appears depends on the parameters of the Lagrangian that defines the background but {\em not} on the effective mass of the dilaton.
\section{Holographic Dense Matter}
Before going to describing the predictions of the new structure uncovered in HLS/VM, let me describe the approach taken from the gravity dual of dense matter in a holographic QCD model. There is a huge literature on the approach to dense and hot matter from the AdS/CFT angle which I cannot possibly review here. Let me focus uniquely on a particular holographic QCD model that is closely related to what is described above in HLS/VM, i.e., the Sakai-Sugimoto model\cite{sakai-sugimoto}. This is the only model that I know of that has the chiral symmetry of QCD in the chiral limit. It describes well -- in the large $N_c$ and large 't Hooft constant $\lambda$ limit where the duality should be reliable and in the probe approximation with $N_f/n_c\ll 1$ -- both meson properties\cite{sakai-sugimoto} and baryon, specially nucleon, properties\cite{hbaryon} which are known to be well described in the quenched approximation in lattice calculations. What is particularly relevant to this discussion is that the model gives a fairly accurate description of nucleon static properties and provides the first ``derivation" of the vector dominance of the nucleon form factor as emphasized in \cite{formfactor}. The latter resolves a long-standing puzzle why Sakurai's vector dominance is strongly violated in the nucleon form factors while it works very well in the meson form factors\footnote{I should point to the remarkable observation by Hashimoto {\it et al.} in \cite{hbaryon} that the form factor of the nucleon vector-dominated by the infinite tower of vector mesons can be re-expressed by a dipole form as found empirically!}.
\subsection{Sakai-Sugimoto model}
The Sakai-Sugimoto action~\cite{sakai-sugimoto} valid in the large $N_c$ and $\lambda$ limit and in the probe (or quenched) approximation can be reduced to the form
\be
S=S_{YM} +S_{CS}\label{SSaction}
\ee
where
\begin{equation}
S_{YM}=-\;\int dx^4 dw
\;\frac{1}{2e^2(w)} \;\tr F_{mn}F^{mn}+\cdots\:,\label{YM}
\end{equation}
with $(a,b)=0,1,2,3,w$ is the 5D YM action coming from DBI action
where the contraction is with respect to the flat metric $dx_\mu dx^\mu+dw^2$. The position-dependent electric coupling  $e(w)$ is given in terms of $N_c$, $\lambda$ and the Kaluza-Klein mass $M_{KK}$ (see Hong et al in \cite{hbaryon})
and
\begin{equation}
S_{CS}=\frac{N_c}{24\pi^2}\int_{4+1}\omega_{5}(A)
\end{equation}
with $d\omega_5(A)=\tr F^3$ is the Chern-Simons action that encodes anomalies. There are two important features in this hQCD model. One is that an action of the same form arises as an ``emergent" gauge action from low-energy chiral Lagrangians with the 5th dimension ``deconstructed."\cite{georgi-theoryspace}. The other is that when viewed in 4D by Kaluza-Klein reduction, it contains an infinite tower of both vector and axial-vector mesons. Now with the low-energy strong interaction dynamics given in the form (\ref{SSaction}), baryons arise as instantons from $S_{YM}$ whose size shrinks to zero as $\sim \lambda^{-1/2}$ as $\lambda\rightarrow \infty$~\cite{hbaryon}. The shrinking to zero size of the instanton is prevented by the Chern-Simons term which takes the form in the presence of the instanton
\be
S_{CS}=\frac{N_c}{8\pi^2}\int {\hat A}_0\tr(F\wedge F)\label{cs}
\ee
where $\hat{A}$ is the $U(1)$ gauge field (analogous to the abelian gauge field, $\omega$, in HLS). The Chern-Simons term provides a Coulomb repulsion that stabilizes the instanton just as the $\omega$ field does to the skyrmion in HLS.

When applied to dense matter, the instanton density distribution $\tr(F\wedge F)$ in (\ref{cs}) is nothing but the baryon number density distribution and hence $\hat{A}_0$ can be interpreted as a variable conjugate to density, i.e., the baryon chemical potential $\mu_B$.

One can have a simple but instructive idea of what this theory contains for dense matter by noting the close parallel between (\ref{lag}) in 4D and (\ref{YM}) in 5D and between (\ref{hwzterm}) in 4D and (\ref{cs}) in 5D. The instanton in (\ref{YM}) stabilized by the $\hat{A}_0$ field in the CS term (\ref{cs}) is an analog of the skyrmion in (\ref{lag}) stabilized by the $\omega$ field in (\ref{hwzterm}). This suggests that if the action (\ref{SSaction}) is treated in mean field for the point-like instantons as baryons balanced by the Coulomb force given by the CS term, there will be  the same defect found with the HLS theory without the dilaton degree of freedom that {\em simulates} the HLS/VM property. One expects that the pion decay constant in this holographic theory will {\em increase} instead of decrease with increasing density. Indeed this was found in a recent analysis of the action (\ref{SSaction}) in cold dense matter\cite{kimetal}.
\subsection{Half-instanton/dyonic phase}
There is at present no quantitative work  in this holographic QCD model of the sort described above for the crystal structure in HLS/VM. But one can reveal a phase that is a gravity dual to the half-skyrmion phase.

Given that the action (\ref{SSaction}) can support instanton solutions for many-baryon systems with the instanton number $A$, the pertinent question one can ask vis-\`a-vis with dense matter is what happens to the instanton at high density. This question was recently answered with the discovery that there can arise a half-instanton matter that is in a dyonic salt form in which chiral symmetry is restored while the baryons are color-singlet hadrons, very much like in the half-skyrmion matter. In fact, the two half-instantons fractionized from one instanton are found to be bound in a bcc configuration, separated by Coulomb repulsion. There is no quantitative description of the dyonic salt configuration at the moment but one can make a plausible qualitative argument using geometry, which, I believe, is robust\cite{RSZ}.

The basic idea is that the flavor instanton in the action (\ref{SSaction}) consists of a superposition of constituents that are BPS dyons in the leading $N_c\lambda$ order. In close analogy to colorons\cite{baal, DIAKONOV} with the colored instanton splitting into constituent dyons as a mechanism for color deconfinement where the Polyakov line plays the central role, here it is also a nontrivial holonomy that works to split the flavor instanton into BPS dyons of opposite charges $e=g=\pm 1/2$. In \cite{RSZ}, geometric reasoning is used for the splitting phenomenon and a dynamical mechanism with topological repulsion balancing the Coulomb forces mediated by the charges of the constituents is invoked for the half-and-half non-BPS structure. It is estimated that the splitting occurs at a density comparable to what was observed in the skyrmion-half-skyrmion transition\cite{park-vento}\footnote{In the skyrmion case, the transition density $n_{1/2}$ was found to be sensitively dependent on the parameters of the Lagrangian. For the parameters fixed in the meson sector, it comes out to be somewhere between $\sim 1.3$ times and $\sim 3$ times the nuclear matter density. This range is roughly what is found in this hQCD model.}. Also the binding energy between the two dyons of opposite charges comes out roughly about 180 MeV, similar to the binding of half-skyrmions in the gauge model\cite{park-vento}.

The arguments leading to the instanton-dyon phase transition is admittedly heuristic and needs to be supported, in the absence of analytical method, by numerical work. But there is no reason to think it is not qualitatively correct. It is easy to understand that with the action (\ref{SSaction}), were  it not for the warping in the conformal direction $w$ in the charge $e(w)$, the constituents of the instanton would be BPS dyons, and hence there would be no reason why the instanton should split into two equal 1/2 instantons. It is the interaction encoded in the curved geometry that turns the BPS dyons with the instanton charge partitioned in $v$ and $1-v$ with $v$ arbitrary into the half-and-half configuration. In this connection, what is highly relevant is the observation made by Hashimoto et al that at short distances, the curvature effect can be ignored and many-body forces between instantons get suppressed\cite{hashimoto}. In the standard nuclear physics picture, short-range repulsion in three-body forces (ignoring four-body forces) is understood to be responsible for pushing phase transitions (e.g., kaon condensation, chiral restoration etc.) in nuclear matter to much higher densities. If holographic duality were to hold in this phenomenon, this feature of many-body forces in the gravity description would imply in the gauge sector that the many-body repulsion problematic in standard nuclear theory is simply absent at high density. This would have an extremely important implication on the EOS of compact stars. In HLS,  many-body forces are also suppressed by the vector manifestation of the vanishing hidden gauge coupling, thus making a valuable prediction that is inaccessible to QCD proper. The two predictions are qualitatively the same although they seem to involve different mechanisms, RG flow matched to QCD for the gauge sector and the infinite tower encoded in the instanton structure in the gravity sector. It would be interesting to ``see" the connection if there is any.
\section{Signals of Half-Skyrmion/Dyonic Phase}
The search for {\em direct} signals of the HLS/VM has been mainly focused on dilepton production in relativistic heavy ion collisions and in EM interactions. The former is for high temperature matter and the latter precision measurement in finite nuclei at nuclear matter density and at zero temperature. The expectation was that via vector dominance, dileptons will be profusely produced from the $\rho$ vector meson in medium with its mass shifted downwards due to the HLS/VM. The searches so far made, however, came out more or less negative, that is, the mass shift associated with the HLS/VM has not been observed. It should, however, be pointed out that this by no means implies that the  HLS/VM prediction is absent in nature. In fact, one of the most remarkable -- and unexpected -- predictions of HLS/VM,  which is found in none of the treatments by the theorists in the field, is that vector dominance is maximally violated as $a\rightarrow 1$ and $g\rightarrow 0$, and the dileptons, ironically, tend to largely decouple from the $\rho$ vector meson, so the observed dileptons carry practically no information on chiral symmetry properties. In fact, what's observed in the $\rho$ spectral function are predominantly mundane nuclear effects due to many-body interactions and do not exhibit clear-cut signals for the chiral structure of the vacuum change that one wants to isolate. The same suppression mechanism takes place both in $T$ and in density. This could indicate  that singling out of a ``smoking-gun" signal for the manifestation of chiral symmetry in dense (and hot) medium at the forthcoming laboratory FAIR/GSI will not be an easy task.

Let me turn to a case which appears to be more promising, i.e., the property of kaons in cold dense matter. Particularly interesting is the behavior of negatively charged kaons (i.e., anti-kaons) fluctuating in the medium described as a dense solitonic background. This concerns two very
important issues in physics of dense hadronic matter. One is the possibility that $K^-$ can trigger strongly correlated mechanisms to
compress hadronic matter in finite nuclei to high density as proposed by Akaishi et al\cite{yamazaki}\footnote{I must mention that there is a considerable controversy on this matter, pros and cons to the Akaishi-Yamazaki idea. For the most recent summary of both theoretical and experimental status, see \cite{gal-weise}. Ultimately experiments will be the jury, and the verdict will have to await the results of those experiments that are being -- and will be -- performed.}. The other issue is the role of kaon condensation in dense compact star matter which has ramifications on the
minimum mass of black holes in the Universe and cosmological natural
selection~\cite{BLR-CNS}. It would be of the greatest theoretical interest to address this problem by means of the instanton
matter given in hQCD, which has the potential to also account for shorter-distance degrees of freedom via an infinite tower of vector
and axial vector mesons. However, no numerical works in this framework are presently available. I shall therefore take the simple Skyrme model with the vector field $V_\mu$ integrated out, implemented with
three important ingredients, viz, the Wess-Zumino term, the soft dilaton field $\chi$ that accounts for the scale symmetry tied to spontaneously
broken chiral symmetry as explained above and the kaon mass term.  Though perhaps oversimplified, I expect it to capture the generic qualitative feature of the novel phenomenon.

The objective is to see how kaons behave on top of the background provided by the skyrmion matter as density is raised. I will assume that the back-reaction of the kaon fluctuation on the background is ignorable. The basic idea is to treat the kaon as ``heavy," not light as in chiral perturbation theory, and wrap it with the skyrmion in the way that Callan and Klebanov did for hyperons\cite{CK}. The appropriate ansatz is \footnote{This ansatz is more convenient for treating kaon fluctuations than the one taken by Callan and Klebanov where the kaon field is sandwiched by the square-root of the soliton field. The two ansatze are found to give an equally good description of the hyperons\cite{riska}.}
\begin{equation}
  U(\vec x,t) = \sqrt{U_K(\vec x,t)} U_0(\vec x) \sqrt{U_K(\vec x,t)},
\label{CKansatz}
\end{equation} where $U_K$ and $U_0$ are, respectively, the fluctuating kaon (doublet) field and the $SU(2)$ soliton field.

\begin{figure}[ht]  
\centerline{\epsfig{file=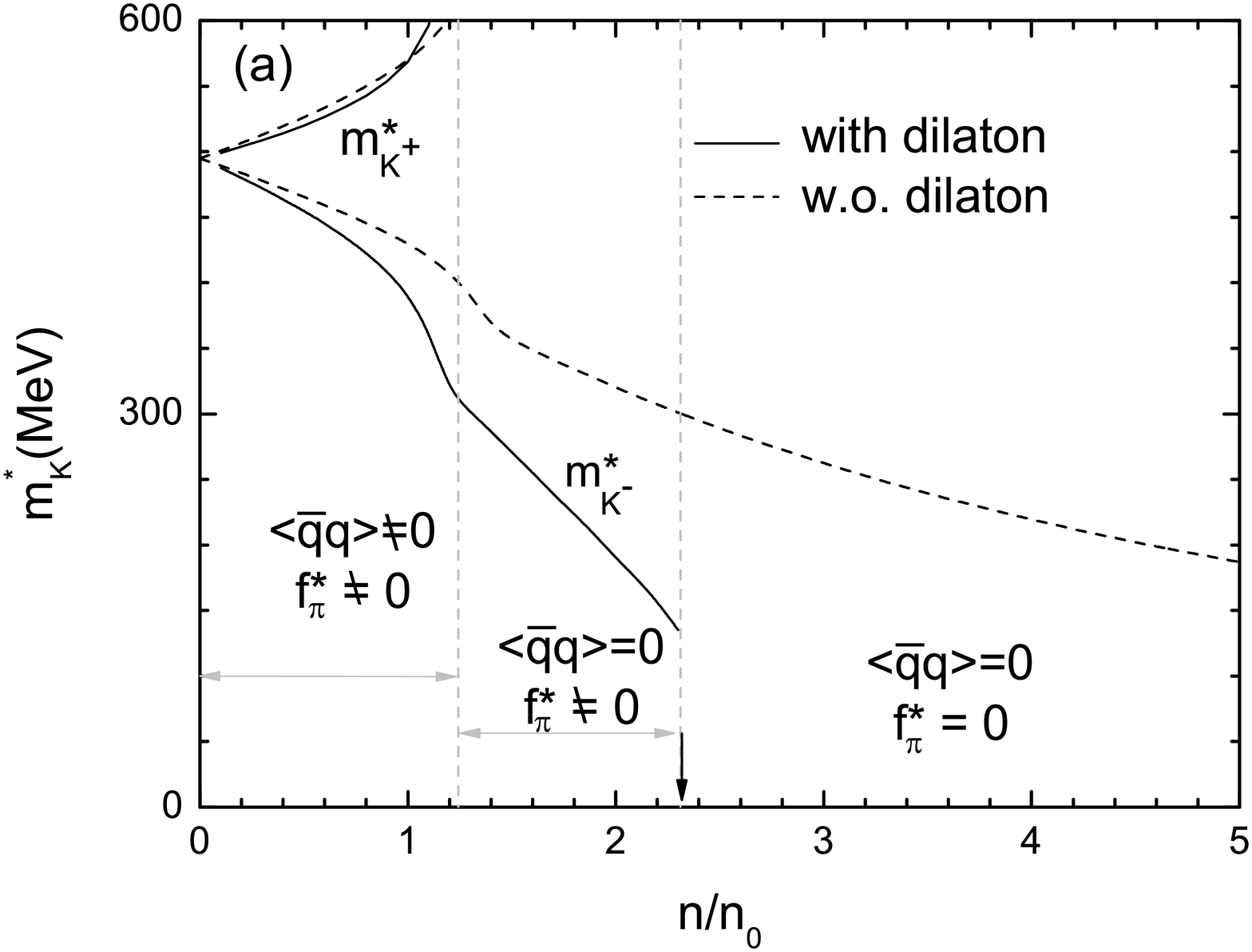, width=6.4cm, angle=0}
\epsfig{file=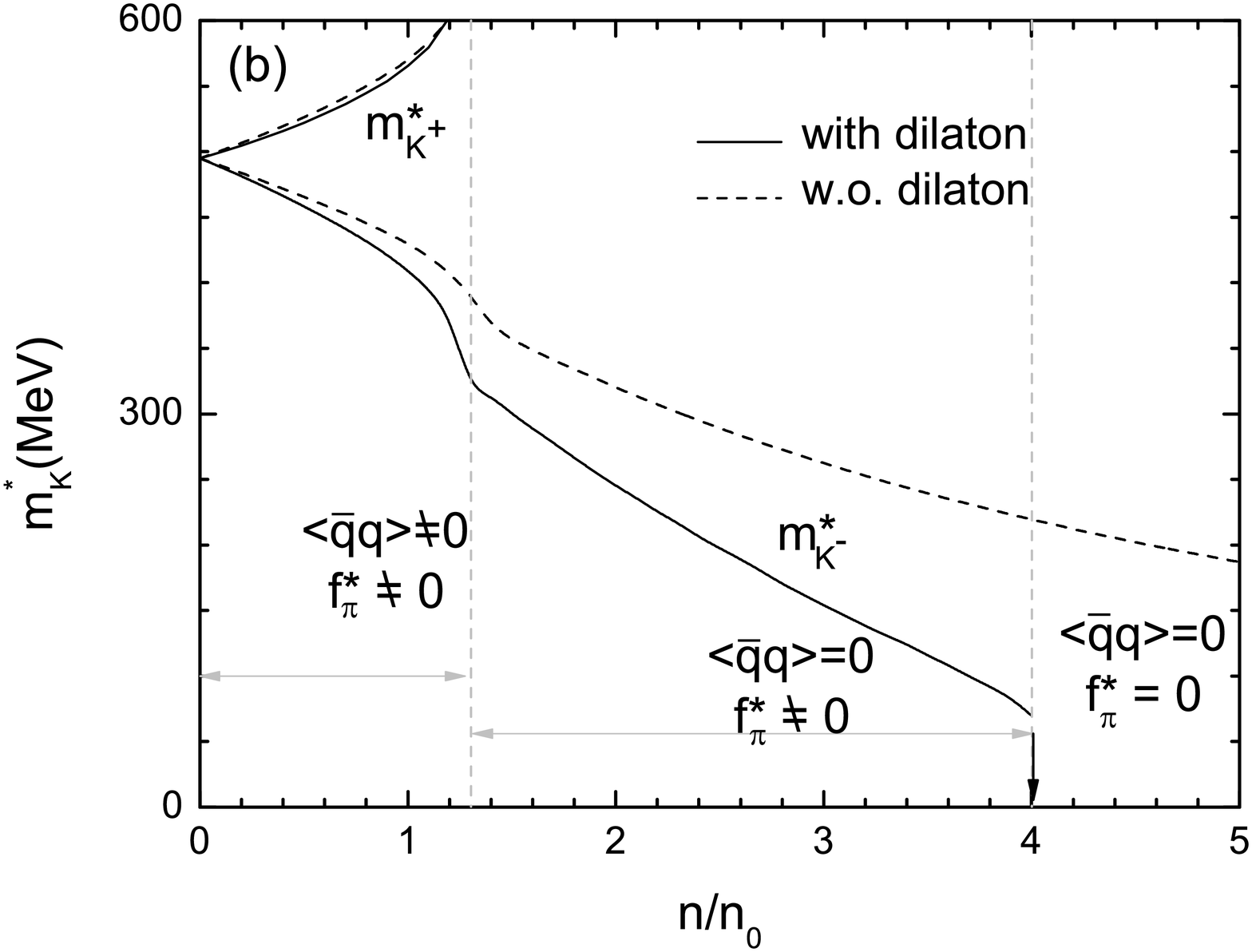, width=6.4cm, angle=0}}
\caption{ $m^*_{K^\pm}$ vs. $n/n_0$ in dense skyrmion matter
that contains three phases as indicated, with the middle being the half-skyrmion phase. The parameters are fixed
at $f_\pi=93$ MeV, $\sqrt{2}ef_\pi=m_\rho=780$ MeV and dilaton mass $m_\chi=600$ MeV
(left panel) and 720 MeV (right pannel). }
\label{fig1}
\end{figure}

The results of this model recently worked out in \cite{PKR} are summarized in Fig.~\ref{fig1}. The mass of the dilaton that figures importantly in this approach is presently unknown. What's taken here are two values, one corresponding to the lowest observed scalar mass $\sim 600$ MeV and the other a value that is used in mean-field theory calculations of nuclear matter using an effective Lagrangian that implements Landau Fermi liquid theory incorporating HLS/VM.  Noteworthy in these results is that up to the threshold of the half-skyrmion phase, the antikaon mass behaves more or less according to chiral perturbation theory\cite{gal-weise}; however once the half-skyrmion phase sets in, the kaon mass drops more steeply, going to zero at about the same density as the chiral transition density $n_\chi\sim (2.3-4)n_0$. This strong attraction that results from the combination of the topological WZ term {\em and} the dilaton field is most probably inaccessible in standard chiral perturbation theory, and hence missing from the chiral perturbation treatments in the literature. This additional attraction could help lead toward the Akaishi-Yamazaki scenario of deeply bound kaonic nuclei. Implementing the mechanism found here in finite nuclei is not an easy task that needs to be worked out.

Another physically interesting question is what does the half-skyrmion
matter do to compact stars? What we have done above is to subject the
kaon to the skyrmion background given in the large $N_c$ limit.
In the large $N_c$ limit, there is no distinction between symmetric matter
and asymmetric matter. Compact stars have neutron excess which typically
engenders repulsion at densities above $n_0$, and hence the asymmetry
effect needs to be taken into account. This effect will arise when the
system is collective-quantized, which gives rise to the leading $1/N_c$
correction to the energy of the bound system. Most of this correction
could be translated into a correction to the effective mass of the kaon, so the rapid dropping of the kaon will have an important impact on the physics of compact stars.
\subsection*{Acknowledgments}
I am grateful for the invitation by Koichi Yamawaki to give this talk. I would also like to acknowledge delightful collaborations with Hyun Kyu Lee, Byung-Yoon Park, Sang-Jin Sin and Ismail Zahed. This work was partly supported by the WCU project of Korean Ministry of Education, Science and Technology (R33-2008-000-10087-0).

\end{document}